\documentclass[a4paper,11pt]{article}

\usepackage{jinstpub}

\usepackage[normalem]{ulem}
\usepackage{comment}
\usepackage{xcolor}
\usepackage{xspace}
\usepackage{caption}
\usepackage{subcaption}

\def\larsoft{\texttt{LArSoft}\xspace}
\def\icc{\texttt{icc}\xspace}
\def\gcc{\texttt{gcc}\xspace}
\def\spack{\texttt{spack}\xspace}

\title{Optimizing the Hit Finding Algorithm for Liquid Argon TPC Neutrino Detectors Using Parallel Architectures}

\author[a]{Sophie Berkman,}\emailAdd{sberkman@fnal.gov}
\author[a]{Giuseppe Cerati,}\emailAdd{cerati@fnal.gov}
\author[a]{Kyle Knoepfel,}
\author[a]{Marc Mengel,}
\author[a]{Allison Reinsvold Hall,}
\author[a]{Michael Wang,}
\affiliation[a]{Fermi National Accelerator Laboratory, Batavia, IL, USA 60510}

\author[b]{Brian Gravelle,}
\author[b]{Boyana Norris.}
\affiliation[b]{University of Oregon, Eugene, OR, USA 97403}

\abstract{
   Neutrinos are particles that interact rarely, so identifying them requires large detectors which produce lots of data. Processing this data with the computing power available is becoming even more difficult as the detectors increase in size to reach their physics goals.  
   Liquid argon time projection chamber (LArTPC) neutrino experiments are expected to grow in the next decade to have 100 times more wires than in currently operating experiments, and modernization of LArTPC reconstruction code, including parallelization both at data- and instruction-level, will help to mitigate this challenge. 
   The LArTPC hit finding algorithm is used across multiple experiments through a common software framework. In this paper we discuss a parallel implementation of this algorithm. Using a standalone setup we find speedup factors of two times from vectorization and 30--100 times from multi-threading on Intel architectures. The new version has been incorporated back into the framework so that it can be used by experiments. On a serial execution, the integrated version is about 10 times faster than the previous one and, once parallelization is enabled, further speedups comparable to the standalone program are achieved.
}


\begin{document}
%
%
\maketitle
\flushbottom
\section{Introduction}
\label{intro}

The US-based neutrino physics program relies on present and future experiments using the Liquid Argon Time Projection Chamber (LArTPC) technology.
The flagship experiment is the future Deep Underground Neutrino Experiment (DUNE)~\cite{Abi:2020wmh}, whose far detector will be made of four liquid argon cryostat modules with 17~kt mass.
It will be located at the Sanford mine in South Dakota and will detect neutrinos from the beam produced 1300~km away at Fermilab.
DUNE has a broad scientific program~\cite{Abi:2020evt} which aims to address fundamental questions about the conservation of the CP symmetry in the lepton sector, the neutrino mass hierarchy, proton decay, and supernovas.
Other LArTPC experiments are already operating or will be turning on in the near future.  The MicroBooNE detector~\cite{Acciarri:2016smi} has a mass $m$ of about 170~t, is located at a distance $L$ of about 400~m from the source of the Booster Neutrino Beam (BNB) at Fermilab, and has been taking data since 2015.
Along the same beamline, ICARUS~\cite{Amerio:2004ze} ($m$=760~t, $L$=500~m) began taking commissioning data in 2020, and SBND~\cite{Admas:2013xka} ($m$=260~t, $L$=100~m) is currently completing the installation phase. 
Together these experiments make up the Fermilab short baseline neutrino program~\cite{Antonello:2015lea}. They aim to understand the low energy electron neutrino anomalies~\cite{Aguilar:2001ty,Aguilar-Arevalo:2018gpe} which are possibly compatible with sterile neutrino oscillations, and to precisely measure neutrino-argon cross sections which are essential for understanding DUNE’s physics results.
In addition, a smaller-scale prototype of the DUNE far detector ($m$=$\sim$700~t), called the ProtoDUNE detector~\cite{Abi:2020mwi}, operated at CERN over the last years and is providing valuable tests of DUNE's detector and reconstruction technologies.  

In LArTPC detectors, neutrinos interact to produce charged particles, these particles lose energy as they travel in the argon and produce ionization electrons.
These drift towards the anode planes of the detector in a homogeneous electric field applied to the argon volume. There are typically three anode planes, each composed of an array of sense wires that measure the ionization charge.
A negative or zero voltage is applied to the first two planes, called induction planes, so that the drifting charge passes through the wires and induces a bipolar signal.
A positive voltage is applied to the last plane, called the collection plane, because the ionization electrons are collected and a uni-polar signal is measured.
Wires are oriented at different angles in the three planes to allow for three-dimensional reconstruction of the neutrino interaction.

Several factors make reconstruction in LArTPC experiments a challenge: the neutrino interaction vertex can be anywhere in the large argon volume, many different topologies can arise from the neutrino interaction, non-negligible electronic noise may overlap with the signal on the wire waveforms, and, for surface detectors, a contamination of tracks from cosmic rays overlaps with the neutrino signal.
Reconstruction is a field of active development for these detectors, and a variety of techniques are being used, from more traditional reconstruction approaches such as Pandora~\cite{Acciarri:2017hat}, to image-based reconstruction using Deep Learning techniques~\cite{Adams:2018bvi}, or tomographic-like paradigms~\cite{Abratenko:2021bzb}.
Early parts of the reconstruction chain, such as wire signal processing and hit finding are more mature however, and in many cases they are shared across otherwise different approaches.

All included, the reconstruction time in a state-of-the-art experiment such as MicroBooNE is on the order of minutes per event. 
The next generation of experiments such as ICARUS and DUNE will produce even more data with 5 and 100 times more wires respectively.  Large increases in reconstruction speed are needed to be able to efficiently process the data from these experiments.
One way to achieve this is with parallel processing of the reconstruction algorithms, which is possible due to the modular structure (cryostats, TPCs, planes, wires) of LArTPC detectors.  This is especially important because trends in computing hardware over the last decades indicate that while increases in clock frequency for single processors will be minimal, large speedups can still be achieved with parallel processing due to the steady increase in the number of processing elements~\cite{comptrends}.

These processing elements can be either vector units within a single core performing the same instruction on different data or multiple cores operating independent instructions. We refer to these two types of parallelism as \emph{vectorization} and \emph{multi-threading}, respectively. 
As demonstrated in the context of other high energy physics experiments (see e.g.~\cite{Lantz:2020yqe}), enabling both types of parallelism is crucial to efficiently utilize current computing resources. Upcoming high-performance computing (HPC) facilities will combine highly parallel CPUs and General Purpose Graphic Processing Units (GPGPUs, or simply GPUs). We note that, while the definitions above apply directly to CPUs, similar concepts are also valid for GPUs; applications to GPUs are being investigated but are beyond the scope of this paper. 
This work demonstrates the speedups that can realistically be achieved with CPU parallelization of LArTPC data processing and the tools to efficiently use these algorithms at HPC centers.

This paper describes the optimization of the LArTPC hit finding algorithm for use on parallel CPU architectures. 
Section~2 introduces the algorithm, Section~3 discusses our optimized implementation in a standalone test program, Section~4 describes its integration into the software framework used by experiments, and Section~5 presents our conclusions.

\section{Hit Finding Algorithm}
The hit finding algorithm is the last in a series of signal processing algorithms that process the electrical signals induced on the detector wires. These signal processing algorithms precede the higher level reconstruction algorithms which directly extract the physics parameters of interest related to the neutrino interaction. The charge on each wire is read out with a particular frequency, which means that the detector signal is a binned waveform with one ADC reading per sampling. 
For example, each of the MicroBooNE detector's 8,256 wires are read out at a frequency of 2 MHz.
The raw signals first go through a noise filtering algorithm which removes excess noise from the waveform signal~\cite{Acciarri:2017sde}. 
The waveforms are then sent through a deconvolution algorithm which corrects for the electric field response and electronic shaping to produce Gaussian-shaped pulsesin waveform intervals identified as "regions of interest" (ROI)~\cite{Adams:2018dra}\cite{Adams:2018gbi}. Each ROI contains at least one of these potential Gaussian pulses, which are called "hits", and are identified by the hit finding algorithm~\cite{Baller:2017ugz}.

\begin{figure*}
\centering
\includegraphics[width=15cm,clip]{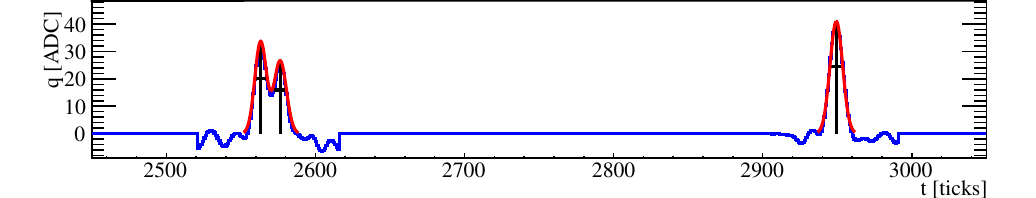}
\caption{Example wire readout (blue) with overlaid hits identified by the hit finding algorithm (red). Pulse height is shown on the y-axis, and time on the x-axis. The region of interest on the right has one identified hit, and the one on the left has two hits. The image is drawn using the \larsoft wire event display~\cite{Snider:2017wjd}.}
\label{fig:wire1}
\end{figure*}

The hit finding algorithm begins with the deconvolved waveform from each wire split into ROIs, and goes through several steps to identify hits. First, the  "CandidateHitFinder" step identifies the separate pulses or groups of pulses within each ROI; for each group, it returns the first and last time tick, as well as the number of Gaussian pulses in the group. There are typically less than five pulses in a ROI.  
Multiple versions of the "CandidateHitFinder" algorithm are available. In all cases they consist of a recursive procedure over waveform intervals, first identifying the leading peak region in the interval, and then repeating the process in the remaining parts of the interval before and after the peak. The main differences between these versions are related to how the peak region is determined, i.e. based on the threshold, derivative, or also applying morphological image processing operations~\cite{Morph} on the waveform.
Then the "PeakFitter" step performs a multi-Gaussian fit in the interval of time ticks of each group, and returns the fit parameters of each Gaussian pulse (peak position, amplitude, and width), as well as the fit $\chi^2$ and number of degrees of freedom. Finally, an optional "HitFilter" step rejects hits that are likely due to noise fluctuations.  The output of the algorithm is a collection of hits on all wires.  Fig.~\ref{fig:wire1} shows an example wire readout zoomed in to areas in which charge is recorded, with identified regions of interest and the fitted Gaussian hits identified in each region.  Each of these steps is configured as a plug-in so that different methods are supported and can be chosen at execution time.  The central version is included in the \larsoft~\cite{Snider:2017wjd} repository, which is a common simulation and reconstruction framework used by all Fermilab LArTPC experiments. 

This algorithm is well suited to demonstrate the potential speedups that may be gained by parallelization of LArTPC reconstruction for several reasons. 
First, this algorithm can be well parallelized because each of the wires and ROIs can be independently processed.  
Also, while this is a simple algorithm, it can take a significant fraction of the reconstruction time, ranging from a few percent to a few tens of percent of the total depending on the experiment. This means that any increases in speed will have a demonstrable impact on the experiment processing times. 

\section{Standalone Hit Finder}

The central \larsoft implementation of the hit finding algorithm was replicated in a standalone piece of code for testing and optimization. 
To study the potential improvements from parallelization, we changed the fitting algorithm from a Gaussian fit routine based on \texttt{Minuit}+\texttt{ROOT}\xspace\cite{Hatlo:2005,Brun:1997pa} to a local implementation of the Levenberg-Marquardt (LM) minimization algorithm~\cite{Bevington:1992, Wang:2018}. 
This choice was motivated by the fact that a custom and lightweight implementation gives us full control of the code, as required for application-specific optimizations.
The LM approach is controlled by a damping parameter $\lambda$ so that, when the fit parameters are far from the minimum, $\lambda$ is large and the algorithm resembles a gradient descent while, when the fit is close to convergence, $\lambda$ becomes smaller and the algorithm approximates
the Gauss-Newton method based on Hessian minimization.
The fit parameters can be restricted to vary only within user-defined boundaries for better fit stability.
Depending on the number of candidate peaks in a ROI, the number of bins in the ROI may vary widely, so the data is more efficiently stored in dynamically-allocated vectors rather than fixed-size arrays.  The overhead due to frequent memory operations is mitigated by the usage of \texttt{jemalloc}~\cite{Evans:2006}.  

Studies of vectorization and multi-threading were done on two computing architectures, a dual processor Intel Xeon Gold 6148 (Skylake, \emph{SKL}) and an Intel Xeon Phi Processor 7230 (Knights Landing, \emph{KNL}).  These were chosen because the SKL processor is relatively mainstream and available, while the KNL processors are currently available at supercomputers.  All KNL results in this section were performed on those available at the Argonne Leadership Computing Facility high-performance computer Theta.  On both SKL and KNL we compiled with \icc~(versions 19.1.1.217 and 19.1.0.166, respectively), as well as with a variety of instruction sets, including AVX-512.  Tests were performed with a sample of 1,000 events of MicroBooNE neutrino simulation overlaid with cosmic muon data, including the real noise and detector response. On average there are 15,600 hits in a MicroBooNE event.

\begin{figure*}
\centering
\includegraphics[width=7cm,clip]{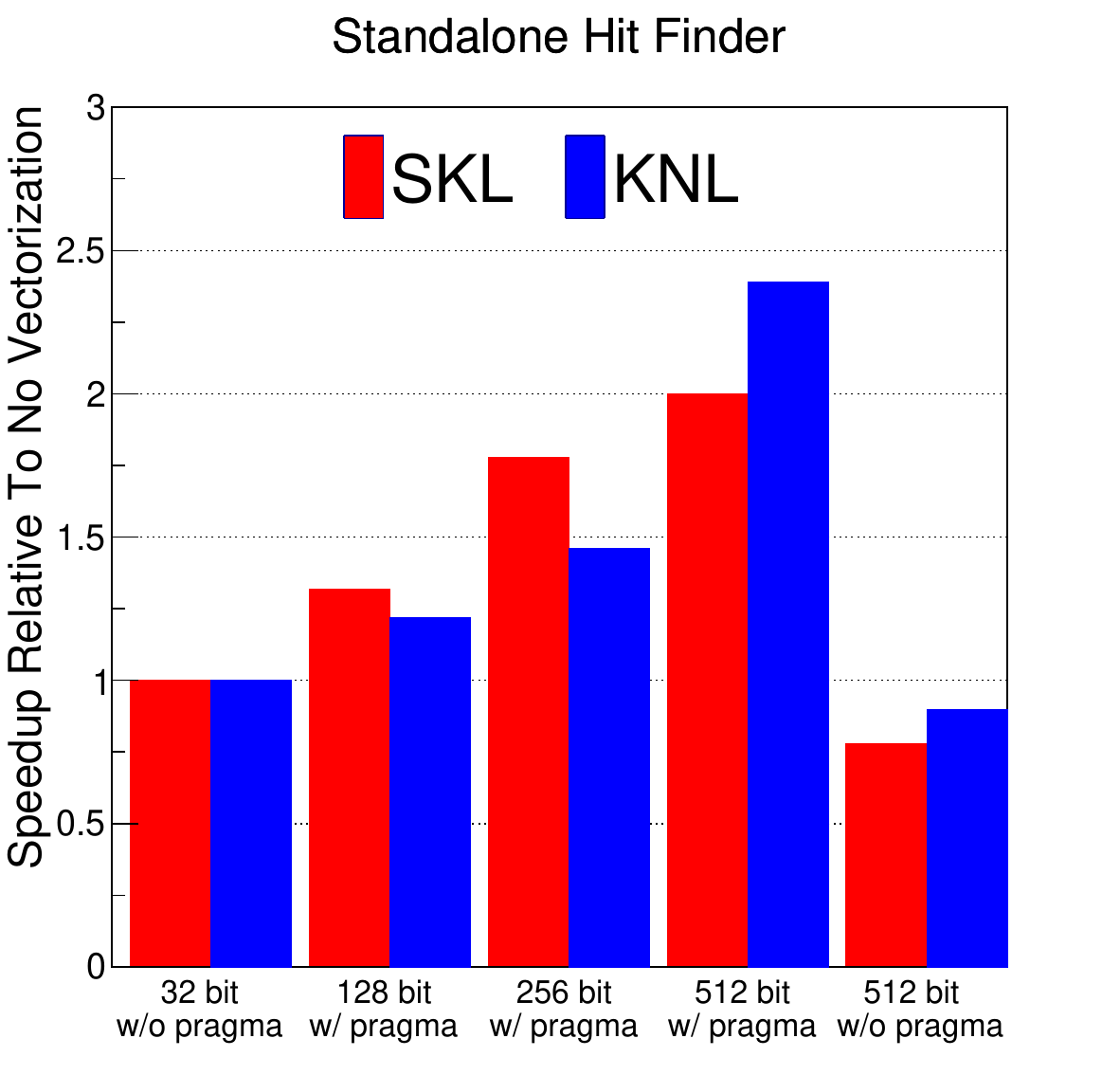}
\caption{Speed increases on SKL and KNL for different compilation configurations. Labels indicate the vector unit utilization in bits specified by different instruction sets, and whether compiler directives (\texttt{pragma}) are used to instruct the compiler on which loops to vectorize.}
\label{fig-mrqdt-vec}
\end{figure*}

We profiled the computational performance of the standalone code with methods such as a roofline~\cite{Williams:rfl} analysis to identify the parts of the code most suited to parallelization.
Results showed that, even with the addition of the LM algorithm, which was found to be about eight times faster than the original algorithm before optimization, 90\% of the time was still spent in the minimization.
One of the challenges of this algorithm is that the number of iterations needed by the fit to converge varies for each hit candidate so it can be difficult to achieve good performance by vectorizing across multiple hit candidates. Therefore, we chose to directly vectorize the most time consuming functions within the LM algorithm. The most important of these functions compute the Gaussian parameters or derivatives, thus performing loops across waveform data bins that can be vectorized.
In order to guide the compiler vectorizing the code, two \texttt{pragma} directives are used: \texttt{pragma simd}, which vectorizes the loop, and \texttt{pragma ivdep}, which instructs the compiler to ignore vector dependencies, as there are not any in the loops that have been vectorized. There are a few limitations to this approach, the first is that only a subset of the code is vectorized.  Another limitation is that the number of bins in an ROI is on the same order as the vector unit size so the relative size of the ``remainder'' part of the loop (defined as $N$ modulo $v$, where $N$ is the number of data bins and $v$ is the vector size) is non-negligible. 
Despite these limitations, we find that both on SKL and KNL processors, when compiling with the AVX-512 instruction set and guiding the compiler with \texttt{pragma} directives, we achieve a factor of two or more increase in speed relative to no vectorization, as shown in Fig.~\ref{fig-mrqdt-vec}.
When the code is compiled including the inter-procedural optimization (IPO) \icc option, the standalone program is $\sim$60\% faster, and the speedups of about two times from  vectorization are preserved; this option does not appear compatible with the \larsoft integration described in the next section, so we show results without it included. 
When the code is compiled on SKL with \gcc~version 10.3.0 we find that, when including \texttt{ffast-math} and \texttt{funroll-loops} options, it has similar performance as the non-vectorized \icc~version. We do not see significant speedups from vectorization with \gcc.

\begin{figure*}
     \centering
     \begin{subfigure}[b]{0.45\textwidth}
         \centering
         \includegraphics[width=\textwidth]{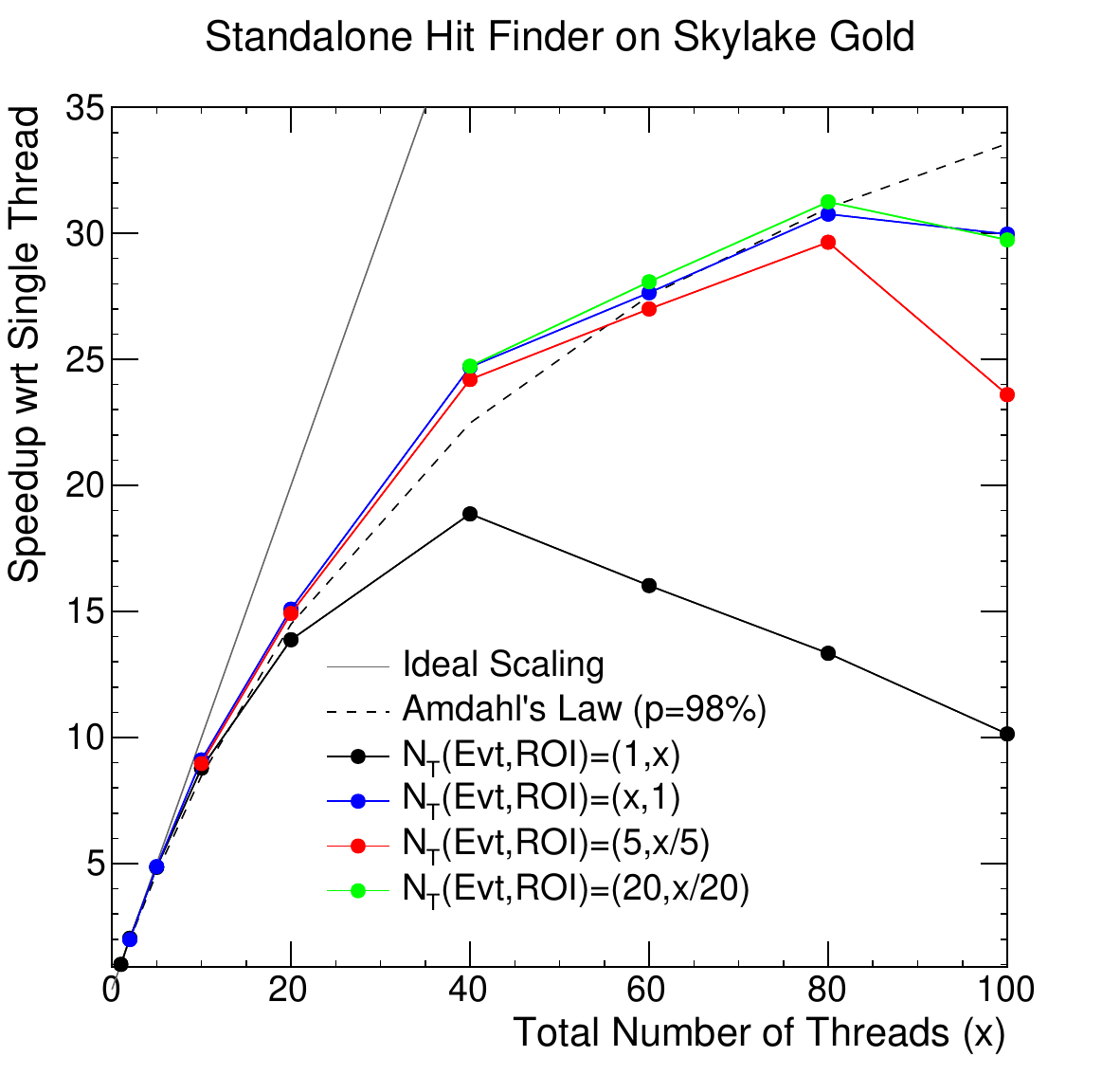}
         \caption{SKL}
    \end{subfigure}
     \begin{subfigure}[b]{0.45\textwidth}
         \centering
         \includegraphics[width=\textwidth]{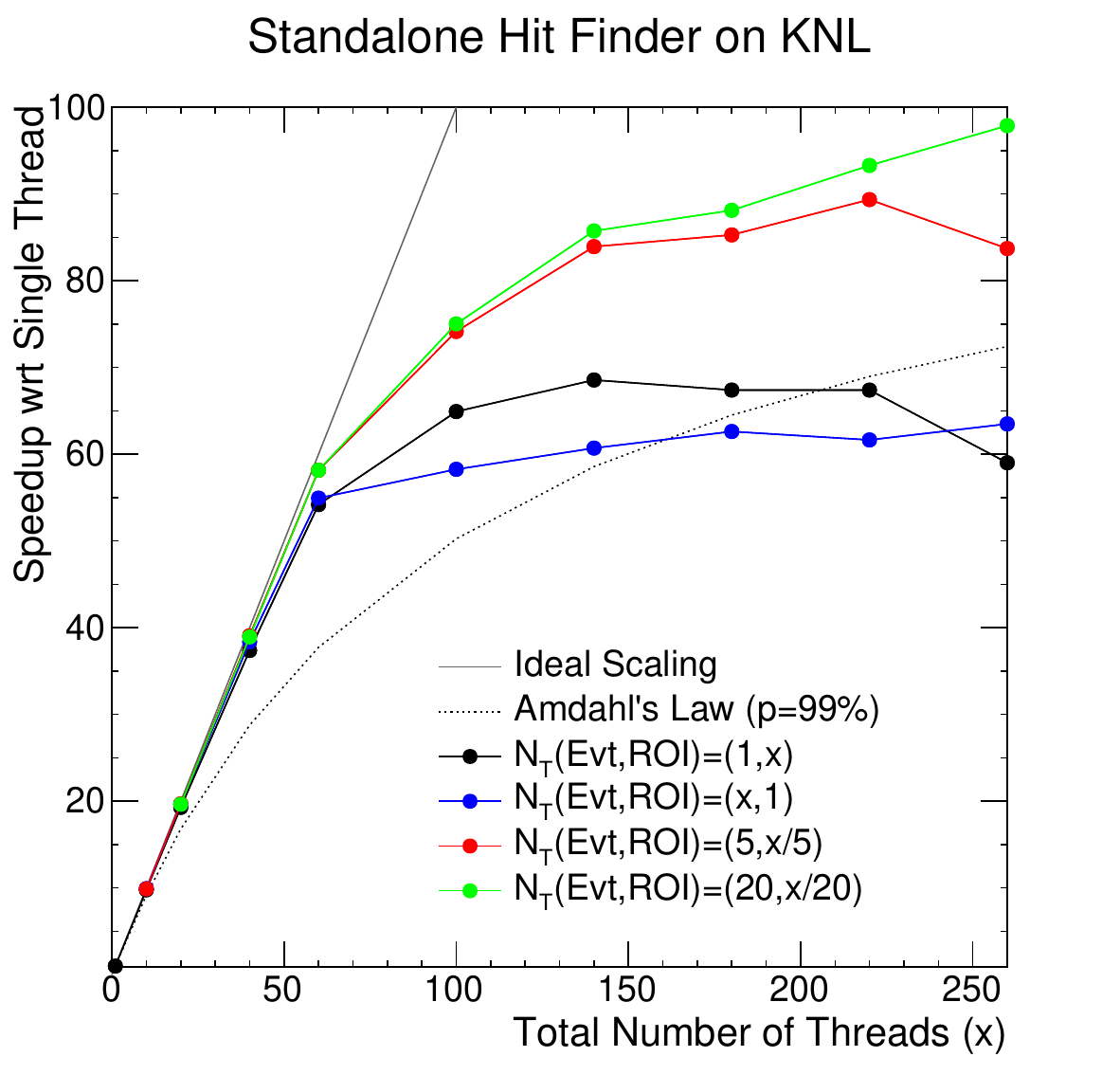}
         \caption{KNL}
         \label{fig:three sin x}
     \end{subfigure}
        \caption{Multi-threading speedups relative to serial execution on SKL and KNL in the standalone hit finder. Each curve indicates a different configuration in terms of number of threads, where each configuration has a fixed number of threads per event or per ROI.  These speedups are also compared to ideal scaling and Amdahl's law. The total number of cores in the SKL (KNL) used is 40 (64).}
        \label{fig-mrqdt-mthr}
\end{figure*}

We implemented multi-threading in the standalone fitting code using \texttt{OpenMP}~\cite{OpenMP}.  There are several levels of parallelism implicit in the signal processing; of particular relevance to this work, the neutrino events are independent, and regions of interest within an event are also independent.  Each of these are looped over separately within the hit finding algorithm, and each of these loops can be directly translated into an \texttt{omp parallel for}, with the ROI loop requiring an \texttt{omp critical} region to synchronize the output of the algorithm. Dynamic thread scheduling along with these two levels of parallelism achieves the best performance.
We find near ideal scaling at low thread counts, and the speed increases by up to a factor of 30 for 80 threads on SKL, and up to a factor of 98 for 260 threads on KNL, as illustrated in Fig.~\ref{fig-mrqdt-mthr}. The scaling performance can be characterized in terms of Amdahl's law~\cite{Amdahl:1967}:
\begin{equation}
    S = \frac{1}{(1-p)+p/N}
\end{equation}
where $S$ is the speedup, $p$ is the parallel fraction of the code, and $N$ is the number of parallel resources used, which is the number of threads in this case. The scaling results of our code correspond to an effective parallel fraction of about 98\% on SKL and better than 99\% on KNL by comparing to the scaling from Amdahl's law. When parallelizing only over regions of interest the parallel fraction is about 97\%.  

In a standalone environment we are able to see significant improvements to the computing time required to identify hits for a LArTPC experiment.  These can then be ported to the experiment code base so that experiments can use them to analyze their data, as will be described in the next section. 

\section{LArSoft Integration}

Standalone programs are useful for development and testing, but in order to make improvements available to experiments they need to be integrated into the central code repository, \larsoft. We completed the integration process in three phases. The goal of the first step was to make the new code readily available to experiments, while the other two focused on enabling parallel execution which is essential for efficient execution at HPC centers.

First, we implemented a new "PeakFitter" plug-in in \larsoft that is based on our LM minimization algorithm. Experiments can use it as part of the hit finding module with a simple change to the configuration file. In this first phase, the execution does not yet leverage parallel computing features. Due to the simplicity of the dedicated minimization algorithm code compared to \texttt{Minuit}+\texttt{ROOT}, however, the speedups are significant. Tests with MicroBooNE, ICARUS, and ProtoDUNE reconstruction workflows using a single thread show the execution time of the full hit finding module in the range of 7-12 times faster, depending on the experiment and the "CandidateHitFinder" plug-in they use. More time-consuming implementations of the "CandidateHitFinder" limit the impact of improvements to the “PeakFitter” when evaluated in terms of speedups of the full hit finder module.

In the second phase of the integration, we enabled multi-threading within the hit finding module. This requires that all components used in the module are thread-safe.  In addition to ensuring thread-safety in our implementation of the "PeakFitter", only minor changes were needed for other components. \larsoft~is based on the \texttt{art} framework~\cite{Green:2012}, which uses \texttt{TBB}~\cite{TBB} for managing multiple thread applications and supports the feature of multi-threaded processing at the event-level. The hit finding module in \larsoft has been adapted to allow multi-threading with nested \texttt{TBB} \texttt{parallel\_for} loops over wires and ROIs, and to make use of \texttt{concurrent\_vector} to store the output of the algorithm in a thread-safe way.
The scaling tests we perform in \larsoft focus on our addition of the sub-event wire level parallelism. Similarly to in the standalone code, the highly variable size of ROIs makes it more efficient to store data in vectors, and \texttt{jemalloc} is used to mitigate the associated overhead and achieve the reported speedups. 
On a SKL machine we find near-ideal thread scaling at low thread counts; speedups are >85\% of the number of threads for thread counts of 5 or below. The maximum speedup is found to be at 60 threads and is 17 times faster than the original.
Overall this corresponds to an effective parallel fraction of 95\% based on Amhdal's law, which is close to the 97\% effective parallelization achieved with ROI-only multi-threading in the standalone application. Second order differences are beyond the scope of this paper and will require further investigation.

The third integration phase targets vectorization speedups. As reported in Section~3, the best performance in terms of vectorization is achieved when compiling the minimization algorithm with \icc and AVX-512. Central \larsoft releases that experiments use in their workflows are typically built with \gcc v8.2.0 using a CMake-based build tool, \texttt{Cetbuildtools}. In order to take full advantage of the vectorization speedups in the minimization algorithm, we compiled \larsoft using \spack as the package manager~\cite{spack, Green:2019}. Both \spack and \icc are typically available at HPC centers and are therefore natural choices when developing workflows targeting these environments. \spack makes it possible to compile the software stack locally on the target architecture and, for each package, allows a custom compiler to be defined with specific options and instruction sets. On a SKL architecture, the hit finder within \larsoft is found to be 2 times faster when the LM minimization algorithm is compiled with \icc and AVX-512 compared to a version compiled with \gcc.  This speedup is consistent with the standalone hit finder result and demonstrates that the implemented methods are a viable way to incorporate vectorization improvements so that they can be used by experiments.

\begin{figure}[h]
\centering
\includegraphics[width=8cm,clip]{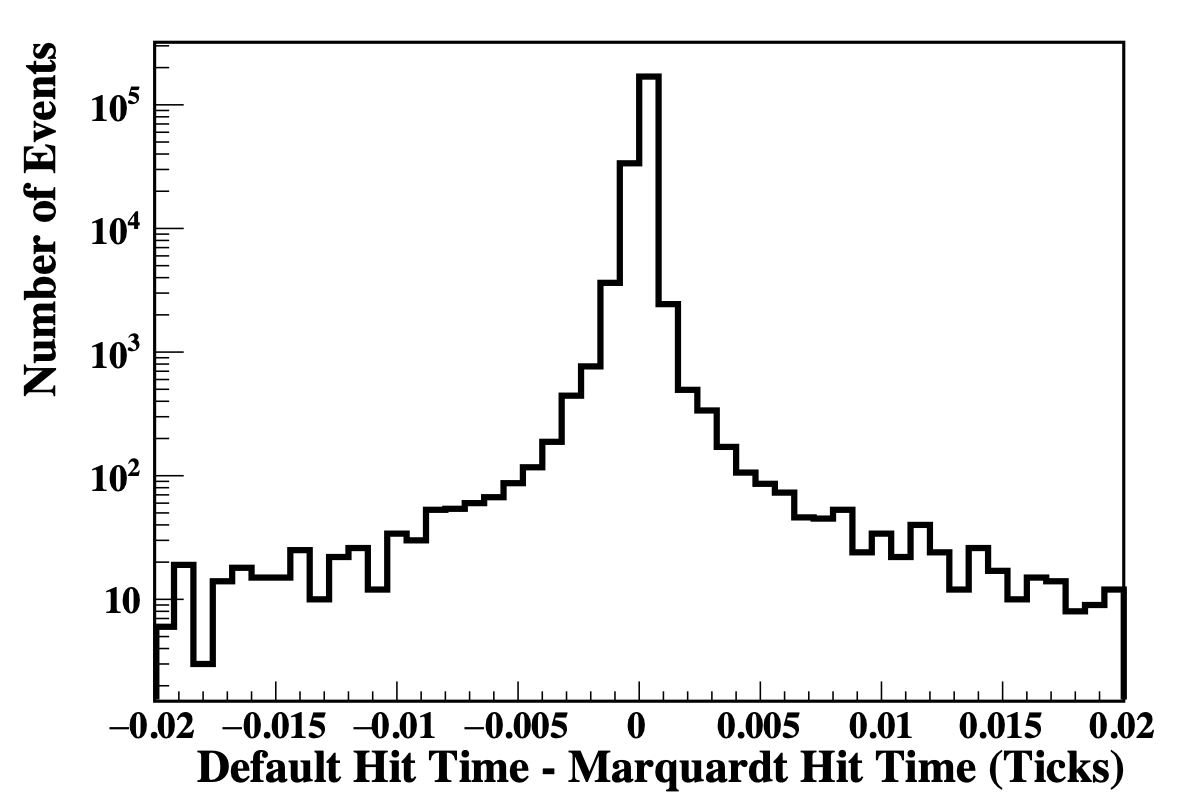}
\caption{Difference in hit peak time when fit with the original and Levenberg-Marquardt based algorithms.}
\label{fig-mrqdt-delta-time}
\end{figure}

We validated the physics output of our algorithm relative to the original by performing a one-to-one comparison of the resulting fit parameters for the same hits. We observe negligible difference between the two, with 98\% of the reconstructed hit times within 0.02 waveform bin units of the original result, as shown in Fig.~\ref{fig-mrqdt-delta-time}. Similar results are obtained when comparing the Gaussian width and amplitude. No loss in terms of efficiency is observed when testing the algorithm across different experiments. Tests have been performed on samples with different noise levels, both from simulation and real data, and confirm the robustness of the algorithm in different experimental conditions. 
The LM minimization algorithm has been adopted in the central reconstruction workflows of the ICARUS and ProtoDUNE experiments.   

\section{Conclusions}

The hit finding algorithm used by liquid argon experiments has been optimized for execution on parallel CPU architectures. This required multi-threading the algorithm at multiple levels and replacing the \texttt{Minuit}+\texttt{ROOT} minimization with a vectorized implementation of the Levenberg-Marquardt method. In a standalone environment vectorization with \icc and the AVX-512 instruction set are found to make the hit finding algorithm twice as fast.  With this setup, multi-threading at the event and sub-event levels led to speedups of up to 100 times on a KNL and 30 times on a SKL.

This version of the hit finding algorithm has been integrated into \larsoft for use by liquid argon experiments. The new algorithm was found to be 7-12 times faster, depending on the experiment, than the original before parallelization, while producing identical physics results.  Vectorization and multi-threading performance within \larsoft is found to be similar to that of the standalone when similar conditions are used in both. 

Looking ahead, this hit finder can be used to improve the time required to process data from liquid argon experiments. 
In addition to decreasing the processing time required in traditional production environments such as on the Open Science Grid~\cite{osg1, osg2}, the tools explored in this work can enable efficient LArTPC data processing workflows at CPU-based HPC facilities, such as ALCF Theta by using all of the available parallel resources, and therefore the HPC systems to their full capacity.  This work can be expanded to apply similar techniques to the optimization of other algorithms.

\section{Acknowledgements}

This material is based upon work supported by the U.S. Department of Energy, Office of Science, Office of Advanced Scientific Computing Research and Office of High Energy Physics, Scientific Discovery through Advanced Computing (SciDAC) program.
Work supported by the Fermi National Accelerator Laboratory, managed and operated by Fermi Research Alliance, LLC under Contract No. DE-AC02-07CH11359 with the U.S. Department of Energy.
This research used resources of the Argonne Leadership Computing Facility, which is a DOE Office of Science User Facility supported under Contract DE-AC02-06CH11357.
We are thankful to the MicroBooNE, ICARUS, and DUNE collaborations for use of their data and simulation samples in this work. 
We are grateful for the support, discussions, and collaboration with the \larsoft and \spack teams from the Fermilab Scientific Computing Division and with the HEP-on-HPC SciDAC project.

\end{document}